# Improving Sample Efficiency in Peptide-HLA Binding Prediction with Hybrid Quantum-Classical Neural Networks

Chenyan Jia[1], Cong Guo[1,*], Siyue Chen[2,3], Pengpeng Ye[2,4], Xiaochun Chen[2,4,*]

[1]Shenzhen SpinQ Technology Co., Ltd., Shenzhen 518048, China
[2]Shenzhen AIage Puhui Longevity Technology Corporation, Ltd., Shenzhen 518048, China
[3]Shien-Ming Wu School of Intelligent Engineering, South China University of Technology, Guangzhou 511442, China
[4]Guangxi Key Laboratory of Longevity Science and Technology, AIage Life Science Corparation Ltd., Nanning 530200, China

[*]Corresponding authors. Cong Guo, E-mail: cguo@spinq.cn; Xiaochun Chen, E-mail: chen.xc@163.com

## Abstract

Peptide-HLA binding prediction is a critical step in neoantigen identification for personalized cancer immunotherapy and holds significant clinical value. However, the training data available for many HLA alleles are extremely limited, which severely constrains the performance of conventional methods on this task. Parameterized quantum circuits are hypothesized to induce inductive biases beneficial for learning from small datasets, yet their application to biological sequence prediction remains underexplored. To address this, we propose a hybrid quantum-classical neural network (HQNN) specifically designed for peptide-HLA binding prediction. HQNN integrates multi-source biological feature encoding with parallel quantum feature extractors and a quantum-enhanced classifier. On two HLA alleles (A*02:01 and B*07:02), HQNN outperforms a parameter-matched classical CNN baseline across all training sizes, with the performance gap widening as training data decreases. Ablation studies confirm the respective contributions of the quantum feature extraction module and the quantum classifier. In noise-aware simulations, performance degrades only mildly, and such degradation is reasonable and acceptable under realistic quantum hardware noise levels. These results suggest that hybrid quantum-classical architectures can provide practical sample-efficiency gains for immunoinformatics tasks in low-data regimes.

**Availability and implementation:** https://github.com/SpinQTech/PepHLA_Interaction.

## 1 Introduction

### 1.1 Background and Clinical Significance

Tumor-specific neoantigen recognition by the adaptive immune system underpins several cancer immunotherapy strategies, including immune checkpoint blockade, adoptive T-cell therapy, and personalized cancer vaccines (Blank et al. 2016; Sahin et al. 2018; Schumacher et al. 2015). Cytotoxic T lymphocytes recognize malignant cells through T-cell receptor interactions with peptide antigens presented by human leukocyte antigen (HLA) molecules on the cell surface (Leone et al. 2013). Peptide-HLA binding represents the first and most selective step in this process, determining which peptides are presented for immune surveillance. Peptides that fail to bind HLA are not presented, regardless of their intrinsic immunogenicity. This selection mechanism makes peptide-HLA binding a primary and tractable target for computational prediction in neoantigen discovery (Yewdell et al. 2022; Rock et al. 2016; Peters et al. 2020). Advances in high-throughput tumor sequencing have enabled the routine identification of large numbers of somatic mutations, generating hundreds to thousands of candidate neoantigens per patient (Mardis 2019; Gubin et al. 2015). However, experimental validation remains costly and time-consuming, necessitating accurate computational prioritization. False positives lead to wasted experimental resources, while false negatives risk missing clinically relevant targets (Borden et al. 2022; Wells et al. 2020). Therefore, improving prediction precision is particularly important for reducing downstream experimental burden. This challenge is further exacerbated by the extensive polymorphism of HLA molecules. Thousands of HLA alleles exist in human populations, each exhibiting distinct peptide-binding preferences. While common alleles are supported by relatively abundant training data, rare alleles often suffer from limited data availability, leading to reduced predictive performance and poor generalization (Sidney et al. 2008; Robinson et al. 2020; Paul et al. 2014; Pearson et al. 2016). This data scarcity issue remains a major bottleneck in immunoinformatics.

### 1.2 Existing Methods and Limitations

A wide range of computational approaches have been developed for peptide-MHC binding prediction over the past two decades. Early methods were primarily based on position-specific scoring matrices (PSSMs) and motif-based models (Parker et al. 1994; Rammensee et al. 1999), followed by classical machine learning approaches such as support vector machines and shallow neural networks (Nielsen et al. 2003; Han and Kim 2017). The NetMHCpan series represents a major advancement, employing ensemble neural networks trained on large-scale binding affinity and eluted ligand data to enable pan-allele prediction (Hoof et al. 2009; Nielsen et al. 2016; Reynisson et al. 2020). More recently, deep learning architectures—including convolutional neural networks, recurrent neural networks, and transformer-based models—have further improved predictive performance, with representative tools such as DeepMHC, MHCflurry, and CAMP (Vang and Xie 2017; Lei et al. 2021; Wang et al. 2023).

Despite these advances, several key challenges remain. First, feature encoding strategies are often fixed or only weakly adaptive. For example, commonly used substitution matrices such as BLOSUM provide predefined representations that may not optimally capture task-specific sequence patterns (Henikoff and Henikoff 1992). Second, existing methods have limited capacity to integrate heterogeneous biological information. Many approaches rely primarily on sequence-based features, while underutilizing complementary information such as structural context, physicochemical properties, and evolutionary signals (Dhusia et al. 2021; Aronson et al. 2022). Third, and most critically, modeling high-order interactions between peptide residues and the polymorphic HLA binding groove remains difficult. These interactions are inherently nonlinear and involve complex dependencies across sequence positions, which are challenging to capture with conventional neural network architectures of practical size. This limitation becomes particularly pronounced in low-data regimes, such as for rare HLA alleles, where insufficient training samples further hinder generalization.

### 1.3 Opportunities in Quantum Machine Learning

Quantum machine learning (QML) has recently emerged as a potential framework for modeling high-dimensional and complex data distributions (Biamonte et al. 2017). By encoding classical data into Hilbert space through parameterized quantum circuits, QML models provide an alternative representation space in which nonlinear feature interactions can be expressed in a high-dimensional manner (Havlíček et al. 2019; Cerezo et al. 2021). Hybrid quantum-classical approaches further combine quantum feature transformations with classical optimization, enabling practical training on near-term quantum simulators and hardware (Mitarai et al. 2018; Huang et al. 2021). One study showed that quantum algorithms demonstrate significant advantages over classical algorithms when learning specific distributed tasks (Lewis et al. 2025). Peptide-HLA binding involves combinatorial interactions across multiple residue positions. Parameterized quantum circuits, through nonlinear transformations and entanglement operations, may provide a flexible mechanism for modeling such complex dependencies in a compact manner. In addition, such representations may introduce an inductive bias that is beneficial in low-data regimes (Caro et al. 2022; Abbas et al. 2021; Alagiyawanna et al. 2024). This property is particularly relevant for peptide-HLA binding prediction, where many HLA alleles are associated with limited training data. While recent studies have explored QML in areas such as protein modeling and drug discovery, its application to immunoinformatics remains relatively underexplored, motivating further investigation.

It should be noted that the objective of this study is not to establish a formal quantum advantage, but rather to empirically investigate whether hybrid quantum-classical architectures can provide practical benefits for modeling peptide-HLA interactions.

### 1.4 Contributions of This Work

To address the limitations of existing approaches, we propose HQNN, a hybrid quantum-classical neural network for peptide-HLA binding prediction. The main contributions of this work are summarized as follows:

First, we introduce a multi-source feature integration framework that jointly encodes sequence-derived and biologically relevant features, including physicochemical properties, secondary structure, intrinsic disorder, and evolutionary information. This design enables more flexible and task-adaptive representations compared to fixed encoding schemes.

Second, we develop a dual-pathway quantum feature extraction module that processes peptide and HLA representations in parallel within a Hilbert space. This module facilitates nonlinear feature transformation and interaction modeling between peptide residues and the HLA binding context.
Third, we construct a hybrid quantum-classical prediction architecture and empirically demonstrate improved performance over a classical counterpart of comparable parameter scale, with more pronounced gains under data-limited conditions. We evaluate HQNN on 9,513 peptide-HLA-A*02:01 samples from the Immune Epitope Database (IEDB) (Vita et al. 2025). Cross-allele validation is conducted on 3,655 peptide-HLA-B*07:02 samples. Experimental results show that HQNN achieves competitive performance compared with existing methods while improving sample efficiency and exhibiting a tendency toward higher precision.

## 2 Methods

### 2.1 Dataset Construction

Experimentally validated peptide-HLA binding data were retrieved from the Immune Epitope Database (IEDB). The dataset was curated using the following criteria: (i) only linear peptides were included; (ii) only MHC ligand binding assays were considered; and (iii) the HLA allele was restricted to HLA-A*02:01. Binding annotations provided by IEDB were used directly, where peptides labeled as binders were treated as positive samples and non-binders as negative samples. After filtering, a total of 9,513 peptide-HLA pairs were retained. The dataset exhibits a positive-to-negative ratio of approximately 1:4, which reflects the inherent class imbalance observed in practical peptide-HLA binding scenarios. The amino acid sequence of HLA-A*02:01 was obtained from UniProt (UniProt Consortium 2025) and used as the reference HLA sequence throughout the study. The dataset was randomly split into training and test subsets, with 80% of the data used for training and 20% reserved for testing. The distribution of peptide lengths across training and test subsets, together with the class composition, is illustrated in Fig. 1. To further evaluate the generalizability of the proposed model across different HLA alleles, an additional HLA-B*07:02 dataset was constructed following the same curation procedure. After filtering, 3,655 peptide-HLA pairs were retained and randomly divided into training (80%) and test (20%) subsets. The resulting dataset exhibited a class distribution comparable to that of HLA-A*02:01, with approximately 20% positive samples in both the training and test sets.

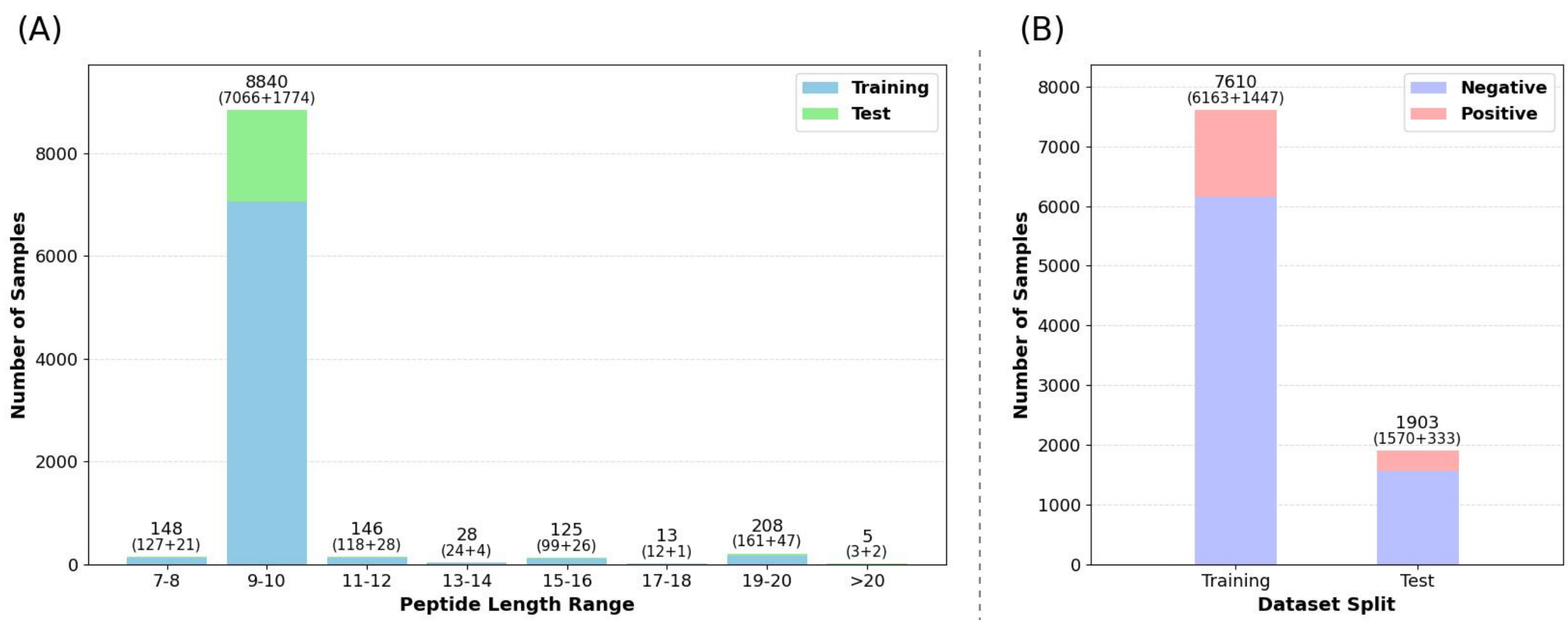

**Fig. 1** Dataset composition and peptide length distribution. (A) Distribution of peptide lengths. (B) Class distribution of positive (binder) and negative (non-binder) samples

### 2.2 Model Architecture

The proposed HQNN framework consists of three components: (i) peptide and HLA feature encoding modules, (ii) peptide and HLA quantum feature extractors, and (iii) a quantum-enhanced classifier. The overall architecture of HQNN is illustrated in Fig. 2. In this framework, heterogeneous biological features derived from peptide and HLA sequences are first integrated into a unified latent representation through classical feature encoding modules. These representations are subsequently mapped to quantum states via amplitude encoding. Two parallel quantum feature extractors are then employed to process peptide and HLA representations independently. Each quantum module applies parameterized quantum circuits to perform nonlinear transformations in a Hilbert space, enabling interaction modeling within transformed feature representations. Finally, the transformed features are combined and fed into a quantum-enhanced classifier, which performs hierarchical feature aggregation and prediction through alternating quantum convolution and pooling operations, followed by classical post-processing layers.

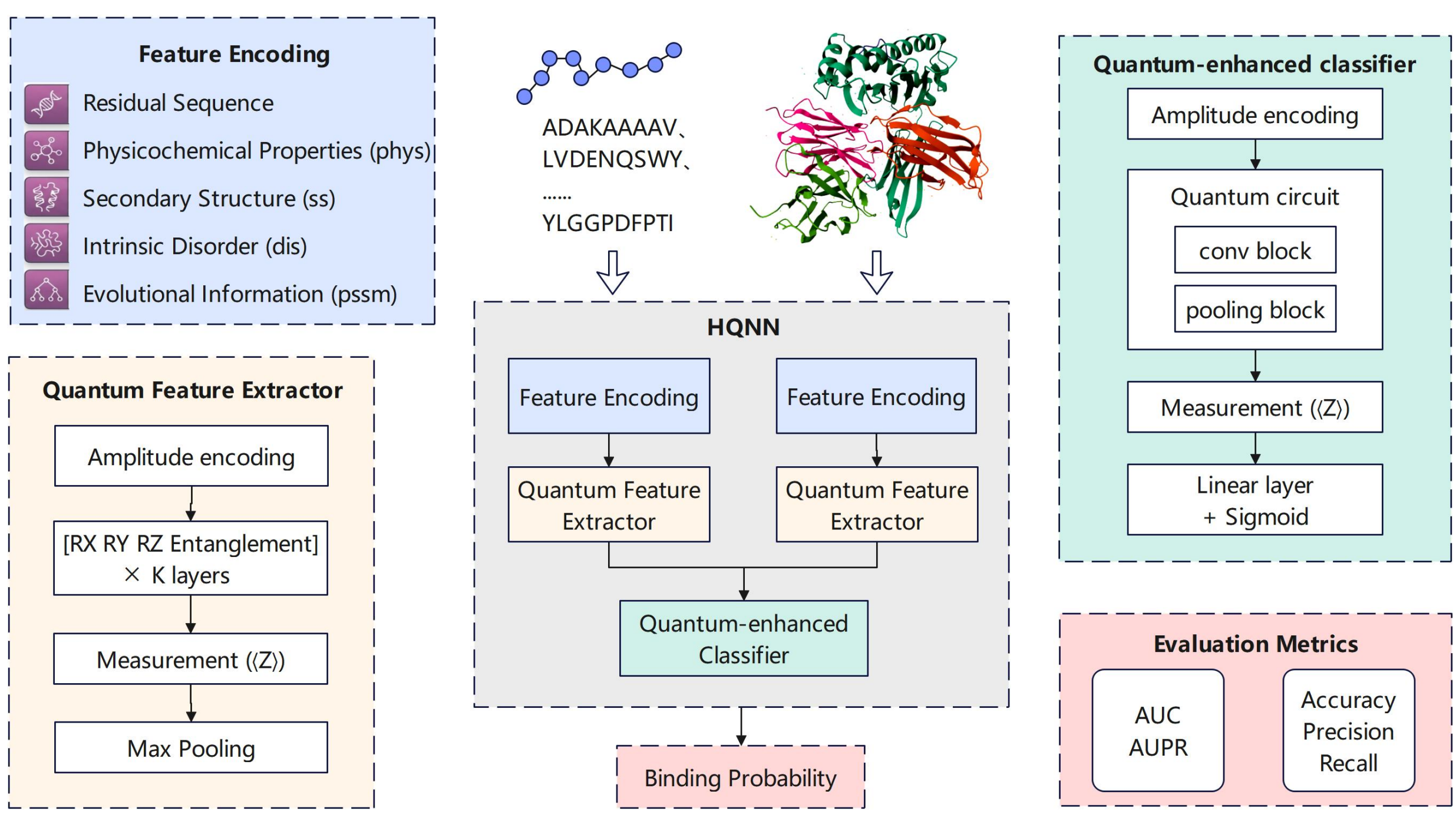

**Fig. 2** Overview of the HQNN framework . HQNN consists of three components: Feature Encoding, Quantum Feature Extractor, and a Quantum-enhanced classifier

## 2.3 Feature Encoding

Formally, given an input sequence $S=\{r_1,r_2,\cdots,r_L\}$, each residue $r_i$ is represented by a composite feature vector constructed from multiple biologically relevant modalities. These modalities are selected to capture complementary aspects of sequence information, including amino acid identity, physicochemical properties, structural context (Magnan and Baldi 2014), intrinsic disorder (Mészáros et al. 2018), and evolutionary conservation (Altschul et al. 1997). To ensure compatibility across heterogeneous feature types, each modality is independently projected into a latent space of equal dimensionality via learnable linear transformations:

$$x_i^{(m)}=W^{(m)}\cdot\varnothing_m(r_i),\quad m\in\{aa,phys, ss, dis, pssm\} \tag{1}$$

where $\varnothing(\cdot)$ denotes feature-specific encoding functions. $W^{(m)}$ represents trainable projection matrices that map each modality into a common embedding space:

$$x_i=Concat(x_i^{(aa)}, x_i^{(phys)}, x_i^{(ss)}, x_i^{(dis)},x_i^{(pssm)}) \tag{2}$$

For peptide sequences, all modalities except evolutionary features are used. For HLA sequences, evolutionary information derived from position-specific scoring matrices (PSSMs) is additionally incorporated. More detailed descriptions of each feature modality and preprocessing procedure are provided in Supplementary Methods S1.

## 2.4 Quantum Feature Extractor

Peptide and HLA sequences are processed using two parallel quantum feature extractors with identical circuit structures. Given a residue-level feature matrix $X\in\mathbb{R}^{L\times d}$, sequences are first normalized to fixed lengths via padding or truncation (peptide $L_p=32$, HLA $L_h=512$). Each residue feature vector $x_i\in\mathbb{R}^d$ is encoded into a quantum state via amplitude encoding. The resulting quantum state is then processed by a parameterized quantum circuit consisting of K layers. Each layer includes single-qubit rotation gates $(R_X,R_Y,R_Z)$ followed by entangling operations implemented via controlled rotation gates $(CRX)$. After quantum evolution, classical features are obtained by measuring the expectation values of Pauli-Z

operators on each qubit:

$$z=(\langle Z_1\rangle, \langle Z_2\rangle, \ldots, \langle Z_n\rangle) \quad (3)$$

where n is the number of qubits. In this study, $n_{pep}$=5 for peptides and $n_{HLA}$=9 for HLA sequences. To obtain sequence-level representations, max pooling is applied along the residue dimension. The outputs of the peptide and HLA branches are then projected into fixed-length embeddings and used for downstream prediction. Further details on circuit design and implementation are provided in Supplementary Methods S2.1. And the visualized quantum circuits for feature extractors are provided in supplementary methods S2.2 and S2.3.

### 2.5 Quantum-enhanced Classifier

The classifier is implemented as a hybrid quantum-classical module composed of multiple hierarchical blocks that alternate between quantum convolution and pooling operations. The quantum convolution operations apply parameterized single-qubit rotation gates followed by entangling gates, enabling local feature transformations across qubits. The pooling operations are designed to progressively reduce the number of active qubits, thereby compressing the feature representation. This is achieved by selectively measuring or discarding a subset of qubits while preserving information in the remaining subsystem, resulting in a reduced-dimensional representation. After passing through the hierarchical blocks, the quantum state is mapped to a low-dimensional representation, from which expectation values of Pauli-Z operators are measured to obtain classical features. These features are then fed into a classical linear layer followed by a sigmoid activation function to produce the final binding probability. Detailed circuit structures for quantum convolution, pooling operations, and the complete classifier are provided in Supplementary Methods S2.4-S2.6.

### 2.6 Evaluation Metrics

Model performance was evaluated using multiple metrics, including the area under the receiver operating characteristic curve (AUC), area under the precision-recall curve (AUPR), accuracy (Acc) , precision (Pre), and recall (Rec). All metrics were evaluated on the held-out test set and reported as mean ± standard deviation (SD) over five independent runs. Training details are provided in Supplementary Methods S2.7. Statistical significance was evaluated using a paired two-sided t-test.

## 3 Results

### 3.1 Sample Efficiency of HQNN in Low-Data Regimes

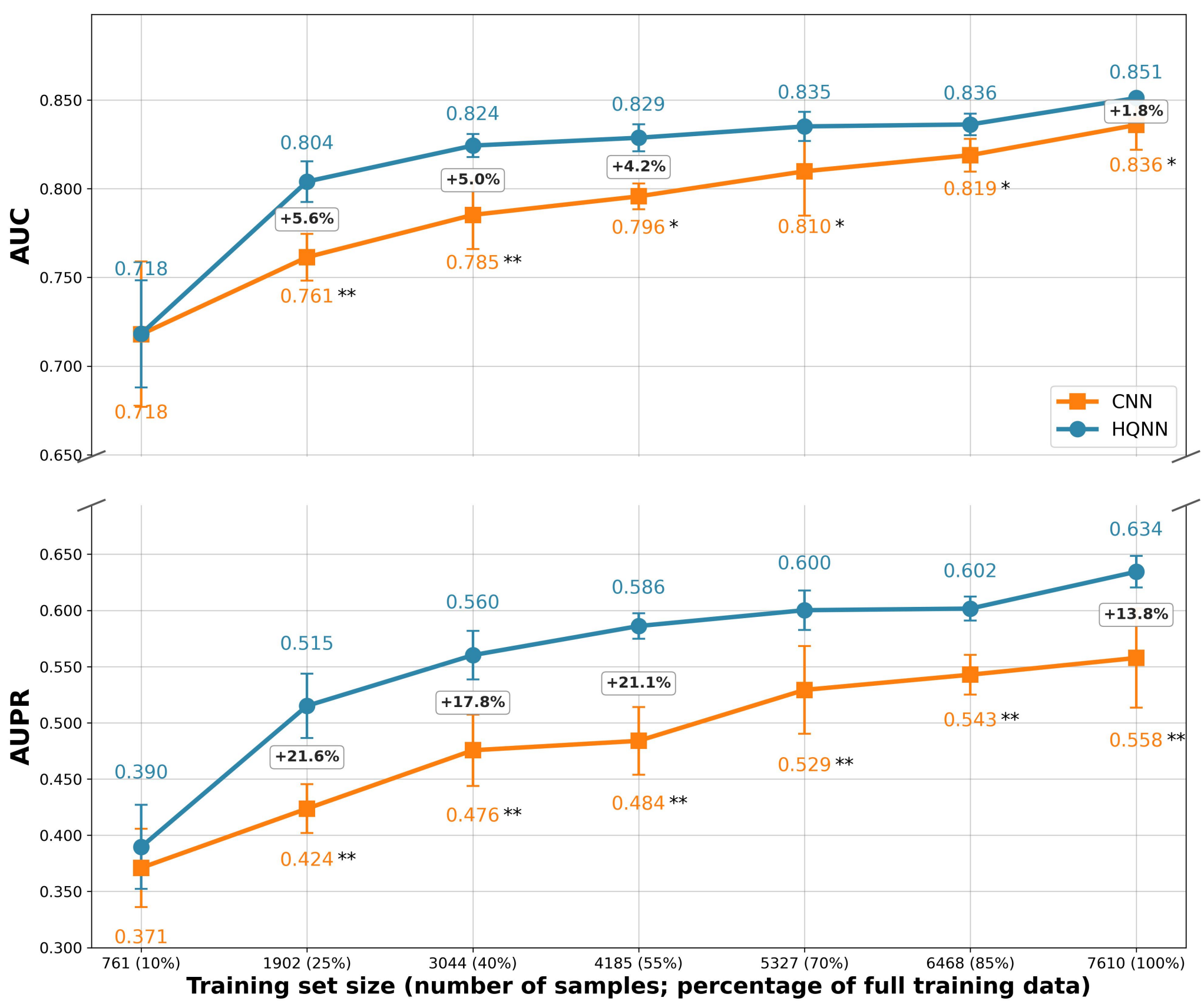

**Fig. 3** Sample efficiency comparison between HQNN and CNN on HLA-A*02:01. Statistical significance uses t-test: *$P$ <.05,

**$P$ <.01

To enable a controlled comparison, we constructed a classical CNN baseline as a direct architectural counterpart of HQNN, in which the quantum components are systematically replaced by classical operations. Specifically, the quantum feature extraction modules are substituted with depthwise separable convolutional layers, and the quantum-enhanced classifier is replaced with fully connected layers. All other components, including input features, preprocessing pipelines, and training protocols, are kept identical. Using this controlled setup, we evaluated model performance under data-limited conditions by randomly subsampling the HLA-A*02:01 training set at proportions of 10%, 25%, 40%, 55%, 70%, 85%, and 100%, while keeping the test set fixed. Fig. 3 reports the AUC for HQNN and the classical CNN baseline. HQNN consistently outperforms CNN across all training sizes, and the performance gap becomes more pronounced as training data decreases. These results indicate that the hybrid quantum-classical architecture learns more informative representations from limited samples, a property particularly valuable for rare HLA alleles. More evaluation metrics and confusion matrices for HLA-A*02:01 training fraction are provided in Supplementary Results. S3.1–S3.2.
To assess the generalization capability of HQNN across different HLA alleles, we conducted additional experiments on the HLA-B*07:02 dataset. As shown in Fig. 4, HQNN consistently outperforms the CNN baseline across all training data scales (10%, 20%, 40%).

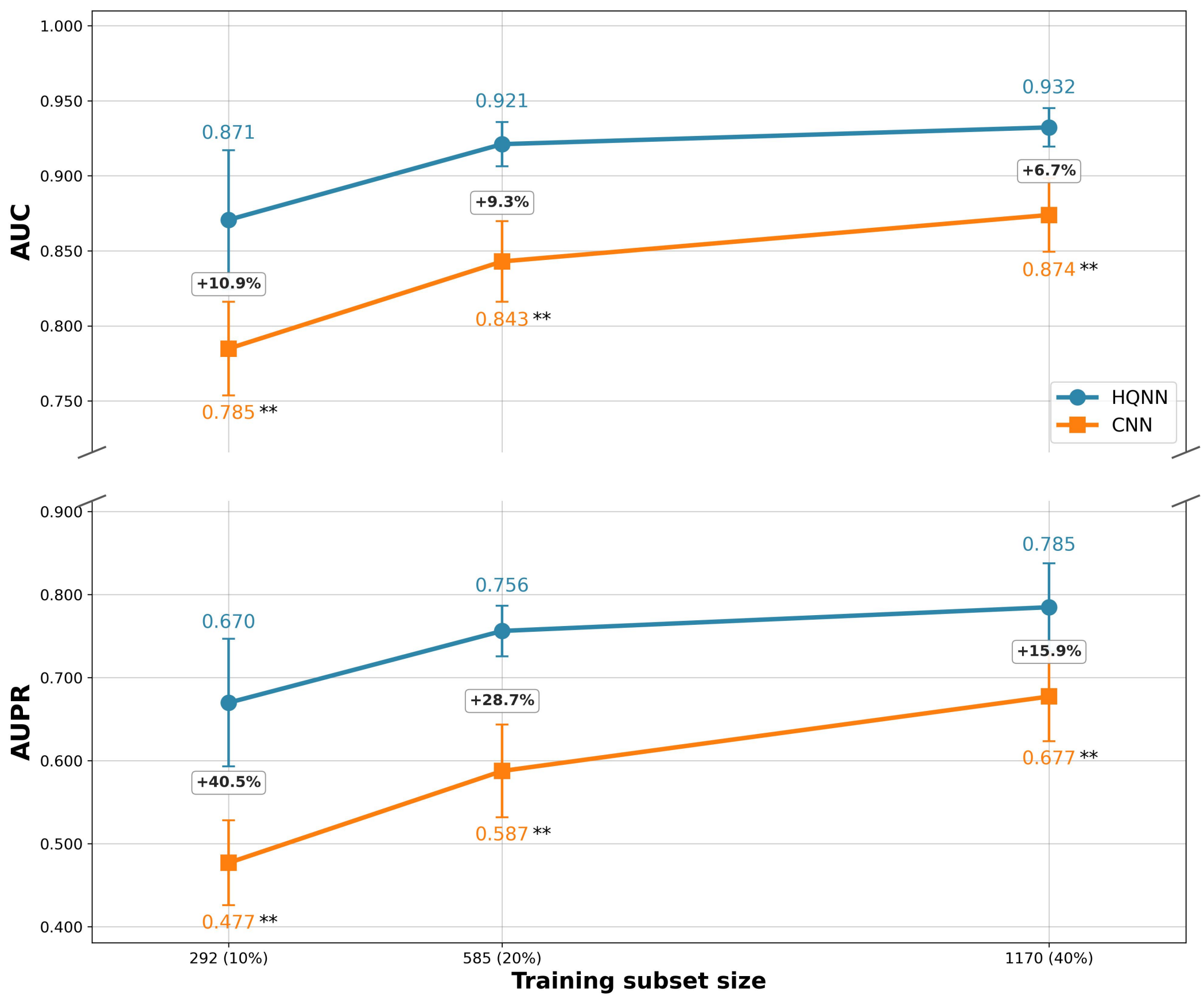

**Fig. 4** Cross-allele performance comparison on HLA-B*07:02 under different training data scales. Statistical significance uses t-test: *$P$ <.05, **$P$ <.01

### 3.2 Impact of Quantum Feature Extractor Depth

To investigate the effect of quantum circuit depth on model performance, we varied the number of variational layers K in the quantum feature extractor while keeping all other components unchanged. The results are presented in Fig. 5. Model performance initially improves as the number of variational layers increases, with both AUC and AUPR showing consistent gains in shallow circuits (K=1-10). However, further increasing circuit depth leads to marginal improvements accompanied by noticeable fluctuations. These results indicate that increasing depth does not continuously translate into better performance and highlight the importance of selecting an appropriate circuit depth. Additional evaluation metrics (accuracy, precision, recall) under varying variational layers are provided in Supplementary Results S3.3.

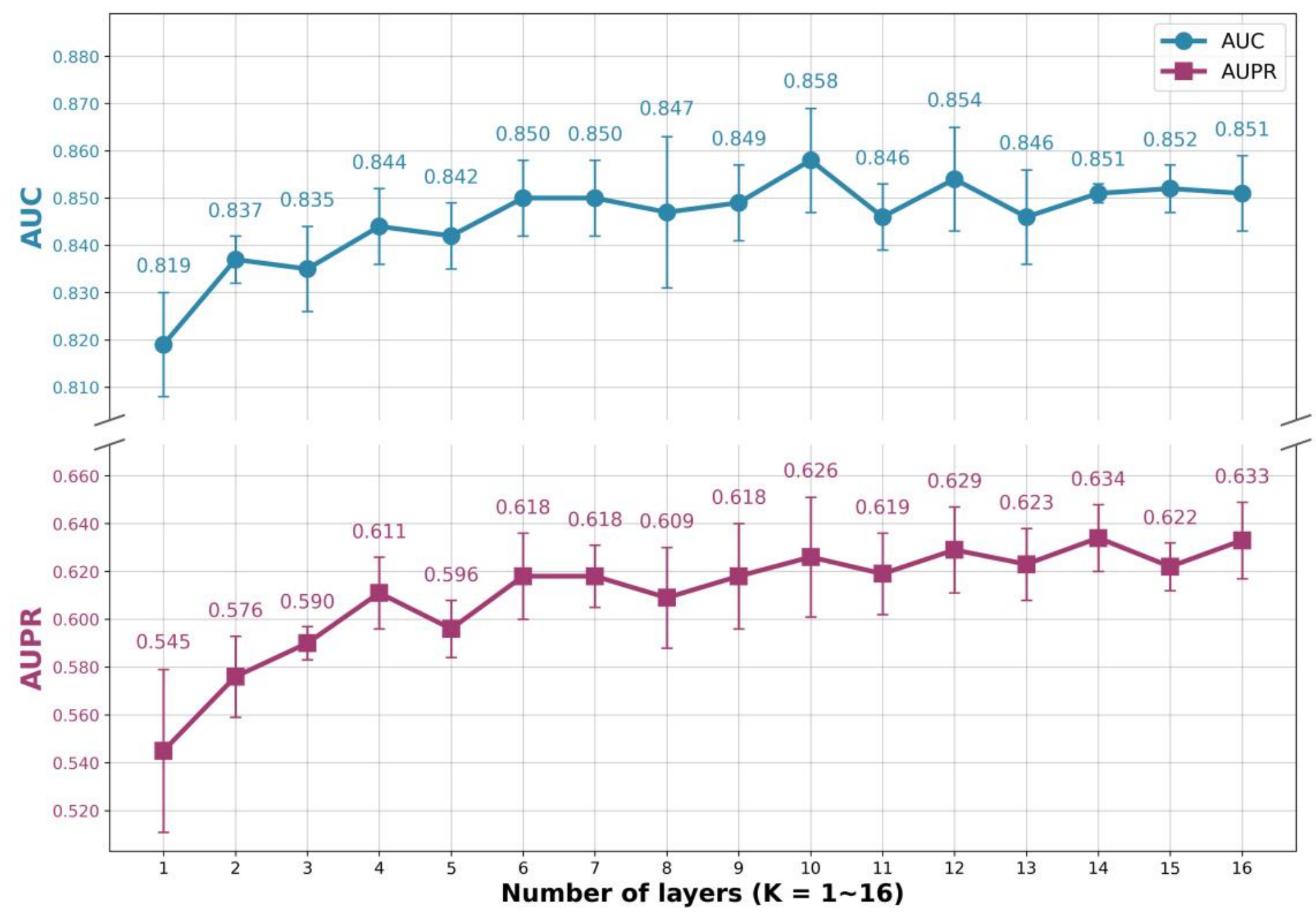

**Fig. 5** Impact of quantum feature extractor depth on model performance

### 3.3 Robustness to Quantum Noise

To assess the robustness of HQNN under quantum noise, we evaluate the model using a noise-aware quantum simulator that incorporates hardware-motivated noise parameters, including relaxation time $T_1$=100μs, dephasing times $T_2$=70μs, and a two-qubit gate error rate of 0.8%. Detailed formulations and implementation are provided in Supplementary Methods S2.8 Noise Modeling. Due to the high computational cost of mixed-state simulation with noise channels, we evaluate the model using quantum extractors with two layers (K=2). The performance comparison is summarized in Table 1.

**Table 1** Performance comparison under ideal and noisy quantum simulation on HLA-A*02:01 binding prediction. Data are reported as mean ± standard deviation (SD) across 5 runs.The best results are **bolded**

| Model | AUC | AUPR | Acc(%) | Pre(%) | Rec(%) |
|---|---|---|---|---|---|
| HQNN (Ideal, K=2) | **0.837** ±0.005 | **0.576** ±0.017 | **85.7** ±0.3 | 63.4 ±1.8 | **43.2** ±3.3 |
| HQNN (Noisy, K=2) | 0.835 ±0.004* | 0.573 ±0.019 | 85.3 ±0.3 | **73.2** ±2.7** | 25.7 ±4.2** |

Paired two-sided t-test: *$P$ <.05, **$P$ <.01.

The AUC decreases only marginally from 0.837 to 0.835, and AUPR shows a similarly small reduction, indicating that the overall ranking capability of the model is largely preserved. Precision increases from 63.4% to 73.2%, while recall decreases from 43.2% to 25.7%. This indicates that the model becomes more selective in assigning positive labels under noisy conditions. From a modeling perspective, this behavior can be attributed to the combined effects of decoherence and stochastic gate errors. In particular, phase damping reduces quantum coherence and suppresses interference patterns, while depolarizing noise introduces random perturbations during entangling operations. Together, these effects reduce the effective expressivity of the quantum circuit, acting as an implicit regularizer that filters out uncertain predictions.

### 3.4 Comparison with Existing Methods

We evaluated the performance of HQNN against three representative approaches: (i) a convolutional neural network (CNN) as a direct counterpart of HQNN, (ii) CAMP (Lei et al. 2021), an advanced deep learning model with substantially larger capacity, and (iii) NetMHCpan4.1 (Reynisson et al. 2020), a widely used method for peptide-HLA binding prediction.
Results are presented in Table 2 and a confusion matrix comparison is provided in Fig. 6. HQNN achieves an AUC of 0.851 ± 0.002 and an AUPR of 0.634 ± 0.014, outperforming the CNN baseline (AUC = 0.836 ± 0.009, AUPR = 0.558 ± 0.018). Given that both models share identical input features, preprocessing pipelines, and training protocols, these results suggest that the introduction of quantum feature transformations provides a more effective representation mechanism under comparable model capacity. In terms of threshold-dependent metrics, HQNN achieves an accuracy of 86.4% ± 0.8, precision of 68.6% ± 1.4, and recall of 49.5% ± 2.6, consistently exceeding the CNN baseline. Notably, the improvement in precision indicates a reduction in false positive predictions, which is particularly desirable in neoantigen screening where experimental validation is costly.

**Table 2** Performance comparison on HLA-A*02:01 binding prediction. The best results are **bolded**. Ratio (n×) indicates that the number of parameters is n times that of HQNN

| Model | Params, Ratio | AUC | AUPR | Acc(%) | Pre(%) | Rec(%) |
|---|---|---|---|---|---|---|
| CNN | 7.1K, 1.1× | 0.836 ±0.009** | 0.558 ±0.018** | 84.9 ±1.4* | 59.3 ±5.9** | 46.4 ±5.0* |
| CAMP-Scr | 3.3M, 508.6× | **0.863** ±0.005* | 0.641 ±0.008 | **86.7** ±1.0 | 65.7 ±2.5* | 51.6 ±3.2 |
| CAMP-FT | 3.3M, 508.6× | 0.849 ±0.004* | 0.631 ±0.008* | 86.1 ±0.6 | 66.2 ±5.4 | 52.1 ±6.4 |
| NetMHCpan4.1 | / | 0.860 | **0.656** | 83.2 | 51.5 | **72.1** |
| HQNN (ours) | 6.5K, 1.0× | 0.851 ±0.002 | 0.634 ±0.014 | 86.4 ±0.8 | **68.6** ±1.4 | 49.5 ±2.6 |

Results are presented as mean ± standard deviation (SD) across 5 runs. Statistical significance was evaluated using a paired two-sided t-test, where $^{*}P<.05$, $^{**}P<.01$; comparisons without annotations are not statistically significant ($P \geq 0.05$). CNN, CAMP-Scr, and HQNN were trained from scratch on our dataset; CAMP-FT was fine-tuned from the pre-trained CAMP; NetMHCpan4.1 results were obtained via its default web interface without allele-specific fine-tuning, and thus serve as a reference rather than a direct comparison.

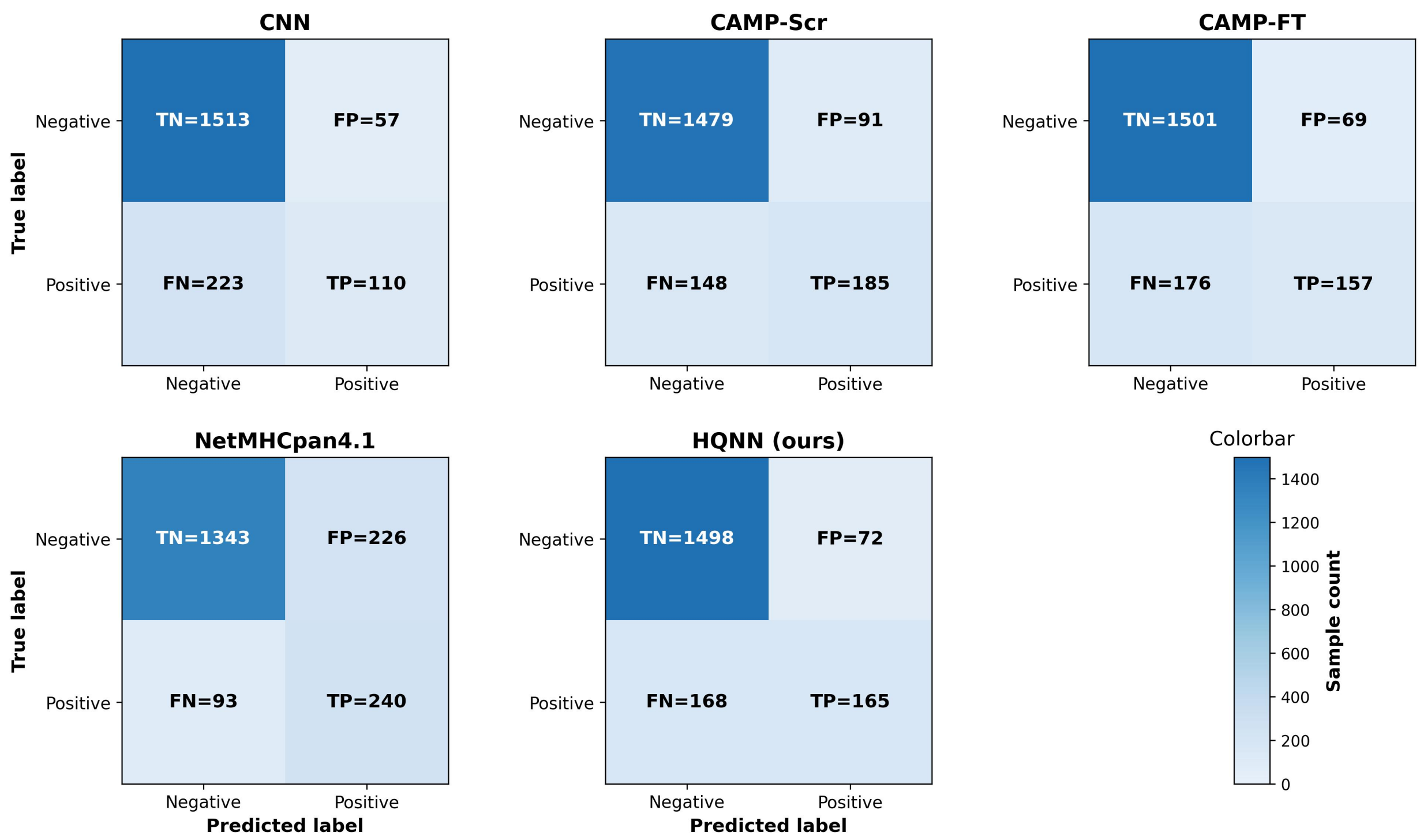

**Fig. 6** Confusion matrix comparison of different methods. Each matrix summarizes prediction outcomes in terms of true negatives (TN), false positives (FP), false negatives (FN), and true positives (TP)

**Table 3** Ablation study on quantum components. The variants are constructed by replacing specific quantum modules with classical counterparts while keeping all other components unchanged.The best results are **bolded**

| Model | AUC (mean ±SD) | AUPR | Acc(%) | Pre(%) | Rec(%) |
|---|---|---|---|---|---|
| HQNN(full) | **0.851** ±0.002 | **0.634** ±0.014 | **86.4** ±0.8 | **68.6** ±1.4 | 49.5 ±2.6 |
| w/o Q-Extractor | 0.836 ±0.010* | 0.588 ±0.021** | 86.2 ±0.5 | 63.1 ±3.1* | **51.4** ±3.5 |
| w/o Q-Classifier | 0.848 ±0.009 | 0.597 ±0.015** | 85.9 ±0.3 | 63.4 ±3.0** | 47.1 ±6.7 |

w/o Q-Extractor : The quantum feature extraction module is replaced with depthwise separable convolution layers.
w/o Q-Classifier : The quantum-enhanced classifier is replaced with fully connected layers.
Paired two-sided t-test: $^{*}P<.05$, $^{**}P<.01$.

### 3.5 Ablation Study

To further investigate the contribution of different quantum components, we conducted an ablation study by selectively removing the quantum components from the full HQNN framework. The results are summarized in Table 3.

Replacing the quantum feature extractor with classical depthwise separable convolution layers leads to a substantial performance degradation, whereas replacing the quantum-enhanced classifier with fully connected layers produces a comparatively smaller effect. These results suggest that the primary contribution of the quantum components arises from feature representation learning rather than the final classification stage.

### 3.6 Feature Space Visualization

To qualitatively assess the representational differences between the classical and quantum-enhanced models, we visualized the peptide-branch features using t-SNE (perplexity = 20). Fig. 7 shows the two-dimensional projections of features extracted by the two models on the test set (333 positive and 333 negative samples). HQNN distinguishes binders from non- binders more clearly: the positive samples form a more compact cluster in the lower-right region. In contrast, the CNN embedding displays a left-right distribution in which positive and negative samples are substantially intermixed.

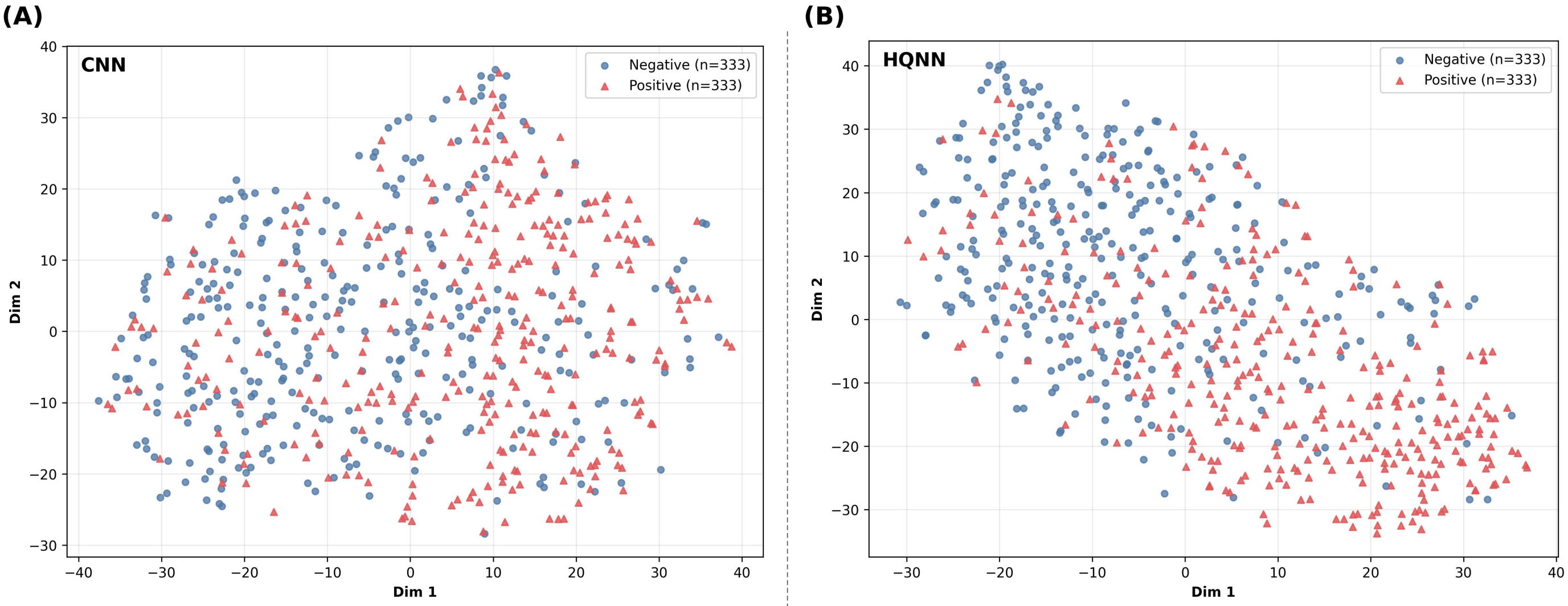

**Fig. 7** t-SNE visualization of peptide features. (A) CNN features. (B) HQNN features. Blue circles represent negative samples (non-binders); red triangles represent positive samples (binders)

### 3.7 Biological Validation: Position-Specific Motif

To evaluate whether HQNN can recover biologically meaningful binding motifs, we performed a position-specific motif recovery analysis for HLA-A*02:01. All peptides in the test set were first filtered to retain only 9-amino-acid sequences to ensure consistent positional alignment across P1-P9. Peptides were ranked according to predicted binding affinity scores, and the top 50 (strict) and top 300 (broad) sequences were selected for further analysis. Heatmaps were generated to visualize residue distributions across the nine peptide positions in Fig. 8.

Both models consistently recovered the canonical HLA-A*02:01 binding motif, with strong positional enrichment at P2 and P9. In particular, leucine (L) was highly dominant at P2, while position P9 showed a preference for valine (V) as the most frequent residue, followed by leucine (L), consistent with established structural knowledge of HLA-A*02:01 binding pockets. Although both models captured these anchor-driven patterns, HQNN exhibited a more concentrated distribution at P2, with leucine enrichment appearing more localized and less dispersed compared to CNN.

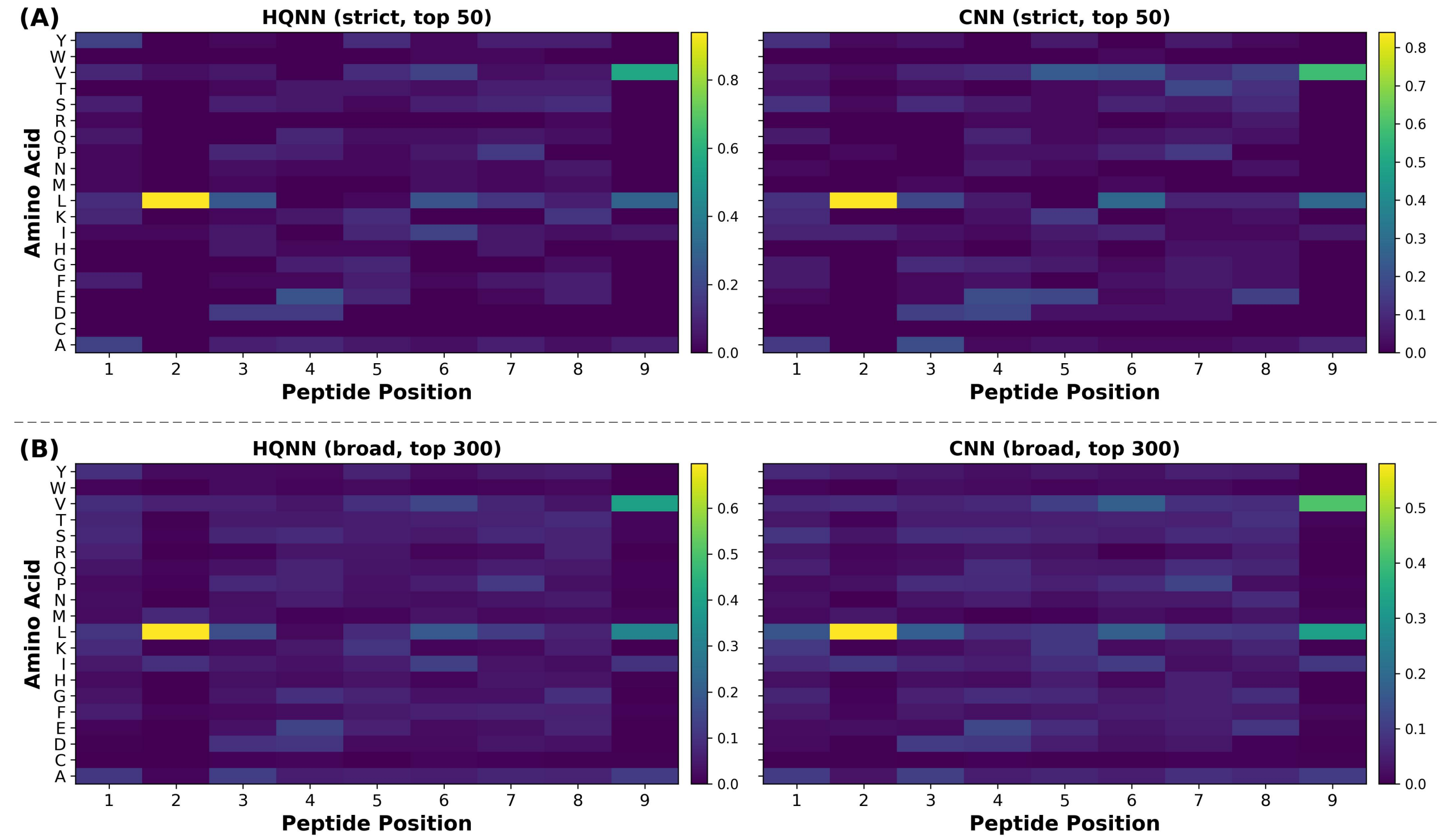

**Fig. 8** Position-specific amino acid frequency heatmaps for HLA-A*02:01 peptide binders. Heatmaps show residue distributions at positions P1-P9 for peptides ranked by predicted binding affinity. Top panels (A) correspond to the top 50 predicted peptides, while bottom panels (B) correspond to the top 300 peptides

# 4 Discussion

## 4.1 Hybrid Quantum–Classical Modeling for Peptide–HLA Binding

HQNN consistently outperformed a parameter-matched classical CNN baseline, with the performance gap becoming more pronounced as training data decreased. This observation suggests that hybrid quantum–classical architectures may provide a beneficial inductive bias for learning peptide–HLA interactions under low-data conditions.

Peptide–HLA binding is governed by complex and position-dependent residue interactions. Parameterized quantum circuits provide an alternative feature transformation mechanism that may capture such nonlinear dependencies more effectively. Consistent with this interpretation, the ablation study shows that removing the quantum feature extractor causes the largest performance degradation, indicating that the primary contribution of the quantum component lies in representation learning. The improved class separation observed in the t-SNE visualization further supports this conclusion.

## 4.2 Biological Relevance

The observed improvement in sample efficiency is particularly relevant for immunoinformatics because many HLA alleles have limited experimentally validated binding data. The ability of HQNN to maintain competitive performance with reduced training samples suggests potential utility for rare alleles and data-scarce scenarios.

In addition, HQNN achieves higher precision than the classical CNN baseline, reducing false positive predictions that require costly experimental validation. Motif recovery analysis further demonstrates that HQNN successfully captures the canonical HLA-A*02:01 anchor preferences at P2 and P9, indicating that the learned representations remain biologically meaningful.

## 4.3 Limitations and Future Directions

Several limitations should be acknowledged. First, evaluation was conducted on only two HLA alleles, and broader validation is needed to assess generalizability across the highly polymorphic HLA system. Second, all experiments were performed on quantum simulators, and future studies should evaluate performance on real quantum hardware. Third, the current architecture models peptide and HLA features independently before fusion. Incorporating explicit residue-level interaction modeling may further improve predictive performance and biological interpretability.

## 5 Conclusion

We proposed HQNN, a hybrid quantum–classical neural network for peptide–HLA binding prediction. Across two HLA alleles, HQNN consistently outperformed a parameter-matched classical CNN baseline, with large gains observed under limited-data conditions. Ablation, feature visualization, and motif recovery analyses suggest that quantum-enhanced feature extraction contributes to improved representation learning while preserving biologically meaningful patterns. These results provide evidence that hybrid quantum–classical architectures can improve sample efficiency in peptide–HLA binding prediction and warrant further investigation for immunoinformatics applications with limited training data.

**Author Contributions** C.J. and C.G. designed the study. C.J. implemented the code, performed the training and evaluation experiments. S.C. assisted with model implementation and experimental execution. P.Y. performed data collection and preprocessing. X.C. provided guidance on biological tools and databases. C.J. wrote the draft manuscript, and C.G. revised it to the final manuscript. All authors, including S.C., P.Y., and X.C., read and approved the final manuscript.

**Funding** This study was supported by Shenzhen Science and Technology Program, China (JCYJ20241202123906009).

**Data Availability** All peptide-HLA binding data were obtained from the Immune Epitope Database (IEDB) at https://www.iedb.org/. The source code, including datasets used in this study, model implementations, pre-trained model checkpoints and evaluation scripts, is publicly available (https://github.com/SpinQTech/PepHLA_Interaction). External tools used for feature generation include IUPred2A (https://iupred2a.elte.hu/), BLAST+ (https://ftp.ncbi.nlm.nih.gov/blast/executables/blast+/LATEST/), and SSPro (https://download.igb.uci.edu/#sspro).

## Declarations

**Competing interests** The authors declare no competing interests.

## References

Blank C U, Haanen J B, Ribas A, et al. The “cancer immunogram”[J]. Science, 2016, 352(6286): 658-660.

Sahin U, Türeci Ö. Personalized vaccines for cancer immunotherapy[J]. Science, 2018, 359(6382): 1355-1360.

Schumacher T N, Schreiber R D. Neoantigens in cancer immunotherapy[J]. Science, 2015, 348(6230): 69-74.

Leone P, Shin E C, Perosa F, et al. MHC class I antigen processing and presenting machinery: organization, function, and defects in tumor cells[J]. Journal of the National Cancer Institute, 2013, 105(16): 1172-1187.

Yewdell J W. MHC class I immunopeptidome: past, present, and future[J]. Molecular & Cellular Proteomics, 2022, 21(7): 100230.

Rock K L, Reits E, Neefjes J. Present yourself! By MHC class I and MHC class II molecules[J]. Trends in immunology, 2016, 37(11): 724-737.

Peters B, Nielsen M, Sette A. T cell epitope predictions[J]. Annual review of immunology, 2020, 38(1): 123-145.

Mardis E R. Neoantigens and genome instability: impact on immunogenomic phenotypes and immunotherapy response[J]. Genome medicine, 2019, 11(1): 71.

Gubin M M, Artyomov M N, Mardis E R, et al. Tumor neoantigens: building a framework for personalized cancer immunotherapy[J]. The Journal of clinical investigation, 2015, 125(9): 3413-3421.

Borden E S, Buetow K H, Wilson M A, et al. Cancer neoantigens: challenges and future directions for prediction, prioritization, and validation[J]. Frontiers in oncology, 2022, 12: 836821.

Wells D K, van Buuren M M, Dang K K, et al. Key parameters of tumor epitope immunogenicity revealed through a consortium approach improve neoantigen prediction[J]. Cell, 2020, 183(3): 818-834. e13.

Sidney J, Peters B, Frahm N, et al. HLA class I supertypes: a revised and updated classification[J]. BMC immunology, 2008, 9(1): 1.

Robinson J, Barker D J, Georgiou X, et al. Ipd-imgt/hla database[J]. Nucleic acids research, 2020, 48(D1): D948-D955.

Paul S, Sidney J, Peters B, et al. Development and validation of a broad scheme for prediction of HLA class II restricted T cell epitopes[C]//Proceedings of the 5th ACM Conference on Bioinformatics, Computational Biology, and Health Informatics. 2014: 733-738.

Pearson H, Daouda T, Granados D P, et al. MHC class I-associated peptides derive from selective regions of the human genome[J]. The Journal of clinical investigation, 2016, 126(12): 4690-4701.

Parker K C, Bednarek M A, Coligan J E. Scheme for ranking potential HLA-A2 binding peptides based on independent binding of individual peptide side-chains[J]. The Journal of Immunology, 1994, 152(1): 163-175.

Rammensee H G, Bachmann J, Emmerich N P N, et al. SYFPEITHI: database for MHC ligands and peptide motifs[J]. Immunogenetics, 1999, 50(3): 213-219.

Nielsen M, Lundegaard C, Worning P, et al. Reliable prediction of T-cell epitopes using neural networks with novel sequence representations[J]. Protein Science, 2003, 12(5): 1007-1017.

Han Y, Kim D. Deep convolutional neural networks for pan-specific peptide-MHC class I binding prediction[J]. BMC bioinformatics, 2017, 18(1): 585.

Hoof I, Peters B, Sidney J, et al. NetMHCpan, a method for MHC class I binding prediction beyond humans[J]. Immunogenetics, 2009, 61(1): 1-13.

Nielsen M, Andreatta M. NetMHCpan-3.0; improved prediction of binding to MHC class I molecules integrating information from multiple receptor and peptide length datasets[J]. Genome medicine, 2016, 8(1): 33.

Reynisson B, Alvarez B, Paul S, et al. NetMHCpan-4.1 and NetMHCIIpan-4.0: improved predictions of MHC antigen presentation by concurrent motif deconvolution and integration of MS MHC eluted ligand data[J]. Nucleic acids research, 2020, 48(W1): W449-W454.

Vang Y S, Xie X. HLA class I binding prediction via convolutional neural networks[J]. Bioinformatics, 2017, 33(17): 2658-2665.

Lei Y, Li S, Liu Z, et al. A deep-learning framework for multi-level peptide-protein interaction prediction[J]. Nature communications, 2021, 12(1): 5465.

Wang M, Kurgan L, Li M. A comprehensive assessment and comparison of tools for HLA class I peptide-binding prediction[J]. Briefings in bioinformatics, 2023, 24(3): bbad150.

Henikoff S, Henikoff J G. Amino acid substitution matrices from protein blocks[J]. Proceedings of the national academy of sciences, 1992, 89(22): 10915-10919.

Dhusia K, Su Z, Wu Y. A structural-based machine learning method to classify binding affinities between TCR and peptide-MHC complexes[J]. Molecular immunology, 2021, 139: 76-86.

Aronson A, Hochner T, Cohen T, et al. Structure modeling and specificity of peptide-MHC class I interactions using geometric deep learning[J]. bioRxiv, 2022: 2022.12. 15.520566.

Biamonte J, Wittek P, Pancotti N, et al. Quantum machine learning[J]. Nature, 2017, 549(7671): 195-202.

Havlíček V, Córcoles A D, Temme K, et al. Supervised learning with quantum-enhanced feature spaces[J]. Nature, 2019, 567(7747): 209-212.

Cerezo M, Arrasmith A, Babbush R, et al. Variational quantum algorithms[J]. Nature Reviews Physics, 2021, 3(9): 625-644.

Mitarai K, Negoro M, Kitagawa M, et al. Quantum circuit learning[J]. Physical Review A, 2018, 98(3): 032309.

Huang H Y, Broughton M, Mohseni M, et al. Power of data in quantum machine learning[J]. Nature communications, 2021, 12(1): 2631.

Lewis L, Gilboa D, McClean J R. Quantum advantage for learning shallow neural networks with natural data distributions[J]. Nature Communications, 2025.

Caro M C, Huang H Y, Cerezo M, et al. Generalization in quantum machine learning from few training data[J]. Nature communications, 2022, 13(1): 4919.

Abbas A, Sutter D, Zoufal C, et al. The power of quantum neural networks[J]. Nature computational science, 2021, 1(6): 403-409.

Vita R, Mahajan S, Overton JA, et al. The Immune Epitope Database (IEDB): 2024 update. Nucleic Acids Research. 2025;53(D1):D444-D452.

Alagiyawanna A, Karunananda A, Mahasinghe A, et al. Enhancing small dataset classification using projected quantum kernels with convolutional neural networks[C]//2024 8th SLAAI International Conference on Artificial Intelligence (SLAAI-ICAI). IEEE, 2024: 1-6.

UniProt Consortium. UniProt: the Universal Protein Knowledgebase in 2025. Nucleic Acids Research. 2025;53(D1):D609-D617.

Magnan C N, Baldi P. SSpro/ACCpro 5: almost perfect prediction of protein secondary structure and relative solvent accessibility using profiles, machine learning and structural similarity[J]. Bioinformatics, 2014, 30(18): 2592-2597.

Mészáros B, Erdős G, Dosztányi Z. IUPred2A: context-dependent prediction of protein disorder as a function of redox state and protein binding[J]. Nucleic acids research, 2018, 46(W1): W329-W337.

Altschul S F, Madden T L, Schäffer A A, et al. Gapped BLAST and PSI-BLAST: a new generation of protein database search programs[J]. Nucleic acids research, 1997, 25(17): 3389-3402.